\documentclass[aps,prb,showpacs,twocolumn,groupedaddress]{revtex4}
\usepackage{amsmath,graphicx}
\usepackage{epsfig}
\usepackage[next]{inputenc}
\begin{document}

%\preprint{Sharif University of Technology}

\title{Thermodynamic Properties of XXZ model in a Transverse Field}

\author{M. Siahatgar}
\affiliation{Physics Department, Sharif University of Technology, Tehran 11155-9161, Iran}
\author{A. Langari}
\affiliation{Physics Department, Sharif University of Technology, Tehran 11155-9161, Iran}
\affiliation{Institute for Studies in Theoretical Physics and Mathematics, Tehran 19395-5531, Iran}
\email[]{langari@sharif.edu}
\homepage[]{http://spin.cscm.ir}

\begin{abstract}
We have numerically studied the thermodynamic properties of the spin $\frac{1}{2}$ XXZ chain in
the presence of a transverse (non commuting) magnetic field. The thermal, field dependence of 
specific heat and correlation functions for chains up to $20$ sites have been calculated. The 
area where the specific heat decays exponentially is considered as a measure of the energy gap. 
We have also obtained the exchange interaction between chains in a bulk material using the random 
phase approximation and derived the phase diagram of the three dimensional material with this 
approximation. The behavior of the structure factor at different momenta verifies the antiferromagnetic
long range order in $y$-direction for the three dimensional case. Moreover, we have concluded that 
the Low Temperature Lanczos results [M. Aichhorn et al., {\it Phys. Rev.  B} {\bf 67}, 161103(R) (2003)]
are more accurate for low temperatures and closer to the full diagonalization ones than the results 
of Finite Temperature Lanczos Method [J. Jaklic and P. Prelovsek, {\it Phys. Rev.  B} {\bf 49}, 5065 (1994)].
\end{abstract}
\date{\today}

%\keywords{}

\maketitle
\section{Introduction}
Quantum phase transition in strongly correlated electron
systems which is the qualitative change in the ground state properties 
versus the a parameter in the Hamiltonian (a magnetic field, amount of disorder, $\cdots$)
have been at the focus of research recently \cite{Sachdevbook,Vojta03}.
Specially, field induced effects in the low dimensional quantum spin models have been attracting 
much interest from theoretical and experimental point of view in recent years
\cite{affleck,uimin,Kenzelmann,Radu,kohgi,dmitriev1,dmitriev2,dmitriev3,langari,gap}.
The magnetic properties of a system with axial anisotropy depends on the direction
of the applied field. For instance, the magnetic properties of the antiferromagnetic 
one dimensional spin 1/2 XXZ chain in the longitudinal field is quite different from
the case of transverse field. 
The longitudinal field commutes with the rest of Hamiltonian and preserves the 
integrability of the model by Bethe ansatz while a transverse field does not commute and
the model is no longer integrable.
The transverse field induces the antiferromagnetic long range
order in perpendicular direction to the field and causes a quantum phase transition to
the paramagnetic phase at the critical point. Experimental observations \cite{Kenzelmann,Radu}
justify this effect in the quasi-one dimensional compound, Cs$_2$CoCl$_4$. 

The qualitative change in the ground state at the quantum critical point (QCP)
for zero temperature ($T=0$)
affects the finite temperature properties of the model close to the QCP.
The quantum critical properties of the Ising model in transverse field has been
extensively studied with different approaches \cite{Sachdevbook,sachdev-young}.
However, there is no work on the finite temperature properties of the XXZ model in the
presence of a transverse field.
It is our aim 
to study the thermodynamic behavior of the anisotropic Heisenberg chain in a transverse field
which is the model Hamiltonian for the mentioned behavior.
The Hamiltonian for this system can be written as
\begin{equation}
\mathcal{H} = J \sum^N_{i=1}(s^x_i s^x_{i+1}+s^y_i s^y_{i+1}+\Delta s^z_i s^z_{i+1} - h s^x_i),
\label{xxztf}
\end{equation}
where $J>0$ is the exchange coupling, $0\leq\Delta\leq1$ is the anisotropy in $z$-direction, 
$h$ is proportional to the transverse magnetic field and $s^{\alpha}_i$ is the $\alpha$-component of
Pauli matrices at site $i$. Experiments\cite{algra} showed that Cs$_2$CoCl$_4$ is an realization of this model with $J=0.23$ meV and $\Delta=0.25$.

We will shortly discuss the different approaches to finite temperature properties of the
lattice model using Lanczos method in the next section. We then implement an appropriate 
approach to get the thermodynamic properties of the XXZ model in the transverse field.
In this paper, we have studied the finite temperature properties of a chain with $N=20$ sites in magnetic fields $h=0 \ldots 5$ and with anisotropies $\Delta=0$ and $\Delta=0.25$.

\section{Finite Temperature Lanczos Method}
The Lanczos diagonalization method is a powerful numerical tool to study the properties of the quantum many body systems on finite clusters. 
It is usually used to get the ground state of a quantum model with high accuracy. However, 
this technique can be extended to finite temperature \cite{jaklic,aichhorn}, where one can study the energy spectrum of the system and its behavior under changing system parameters such as anisotropy and magnetic field. 

In the Lanczos algorithm, the ground state can be obtained with very high accuracy. However, to get finite 
temperature behavior we need to take average over the whole Hilbert space ($L$) with Boltzmann weights, i.e.

\begin{equation}
\langle \mathcal{O} \rangle = \dfrac{1}{\mathcal{Z}} \sum^L_n \langle n | \mathcal{O} e^{-\beta \mathcal{H}} | n \rangle, \quad \mathcal{Z} = \sum^L_n \langle n | e^{-\beta \mathcal{H}} | n \rangle
\end{equation}

It has been proposed that the Lanczos procedure can be used to get reliable approximation
for the finite temperature ($T\neq0$) properties of the lattice models \cite{jaklic}. In this approach
the Hilbert space is partially spanned by different random initial vectors for several Lanczos
procedures. This approach is called Finite Temperature Lanczos Method (FTLM)\cite{jaklic}
which is based on the following equations.
\begin{align}
\langle \mathcal{O} \rangle &\simeq \dfrac{1}{\mathcal{Z}} \sum^R_r \sum^M_m  e^{-\beta \varepsilon^{(r)}_{m}} \langle r | \Psi^{(r)}_m \rangle \langle \Psi^{(r)}_m | \mathcal{O} | r \rangle \\ 
\mathcal{Z} &\simeq \sum^R_r \sum^M_m  e^{-\beta \varepsilon^{(r)}_{m}} {\big| \langle r | \Psi^{(r)}_m \rangle \big|}^2,
\label{ftlm}
\end{align}
where $\beta=1/k_BT$ ($k_B=1$) is the Boltzmann constant and $R\sim 10$ is the total number of random samples with different initial vectors $| r \rangle$. The eigenvectors and the
corresponding eigenvalues of the tridiagonal matrix are $| \Psi^{(r)}_m \rangle$ and
$\varepsilon^{(r)}_{m}$ respectively for $m=1,\ldots, M$. The FTLM procedure works well 
for finite temperatures, however, it does not converge to the ground state expectation value
($\langle \psi_0 |\mathcal{O}|\psi_0 \rangle$) as $T\rightarrow 0$ because of statistical fluctuations. i.e. 

\begin{equation}
\langle \mathcal{O} \rangle = \sum^R_r \langle \Psi_0 | \mathcal{O} | r \rangle \langle r | \Psi_0 \rangle /
 \sum^R_r \langle \Psi_0 | r \rangle \langle r | \Psi_0 \rangle 
\end{equation}
This problem can be solved by a symmetric algorithm which is explained below.

\subsection{Low Temperature Lanczos Method}
The Low Temperature Lanczos Method (LTLM) was proposed as a symmetric algorithm to
remedy failure of FTLM at low temperatures\cite{aichhorn}. However, LTLM needs more CPU time and memory but
requires fewer steps in low temperatures \cite{siahatgar}. LTLM can be formulated similar to the case of FLTM
with the following equation,
\begin{equation}
\begin{split}
\langle \mathcal{O} \rangle \simeq\dfrac{1}{\mathcal{Z}} \sum^R_r \sum^M_{i,j}  e^{-\frac{1}{2}\beta (\varepsilon^{(r)}_{i}+\varepsilon^{(r)}_{j})} \times &\\ \langle r | \Psi^{(r)}_i \rangle \langle \Psi^{(r)}_i | \mathcal{O} | \Psi^{(r)}_j \rangle \langle \Psi^{(r)}_j | r \rangle,
\end{split}
\label{ltlm}
\end{equation}
where the partition function ($\mathcal{Z}$) is calculated in the same way as in Eq. (\ref{ftlm}). Using this method, as $T\rightarrow 0$ we have,

\begin{equation}
\langle \mathcal{O} \rangle = \sum^R_r \langle \Psi_0 | r \rangle \langle r | \Psi_0 \rangle \langle \Psi_0 | \mathcal{O} | \Psi_0 \rangle / \sum^R_r \langle \Psi_0 | r \rangle \langle r | \Psi_0 \rangle  
\end{equation}

which is equal to the ground state expectation value. It is thus more accurate to get the
low temperature quantum properties.

\section{Finite Temperature Results}

\subsection{Comparison between LTLM and FTLM}
\begin{figure}
\includegraphics[width=8cm]{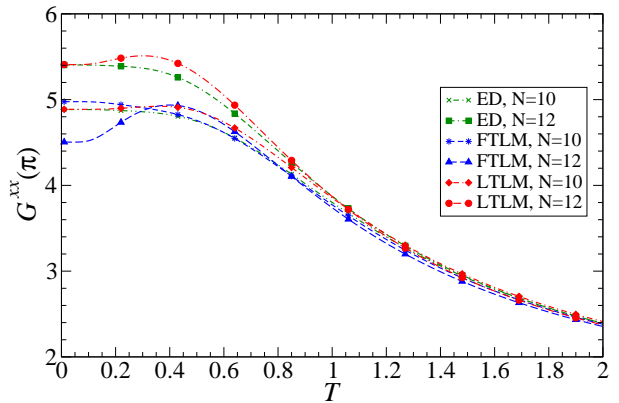}
\caption{The $x$-component spin structure factor at momentum $\pi$ versus temperature
for a chain of length $N=10, 12$ using FTLM, LTLM and exact diagonalization (ED).(Color online)}
\label{fig1}
\end{figure}

We are interested in the finite temperature properties of the Hamiltonian defined
in Eq. (\ref{xxztf}) and specially in the quantum critical behavior as $T\rightarrow 0$.
We have thus implemented the LTLM  method to get the finite temperature properties which
is accurate close to $T=0$, although it consumes more CPU time and memory. To justify this, 
we have plotted the $x$-component spin structure factor at momentum $\pi$ 
($G^{xx}(\pi)$, see Eq. (\ref{structurefactor}))
versus $T$ in Fig. \ref{fig1}. In this figure, we have shown both the FTLM and LTLM results
for a chain of length $N=10, 12$ together with the exact diagonalization (ED) results for 
comparison. The discrepancy of FTLM in low temperature ($T \lesssim 0.2$) in comparison with
the ED results is clear specially for $N=12$ as discussed in previous section. Fig. \ref{fig1} shows that the LTLM results
converge to the ED ones as $T\rightarrow 0$ and are in good agreement for higher temperatures.

\subsection{Specific Heat}
\begin{figure}[ht!]
\begin{center}
\includegraphics[width=8cm]{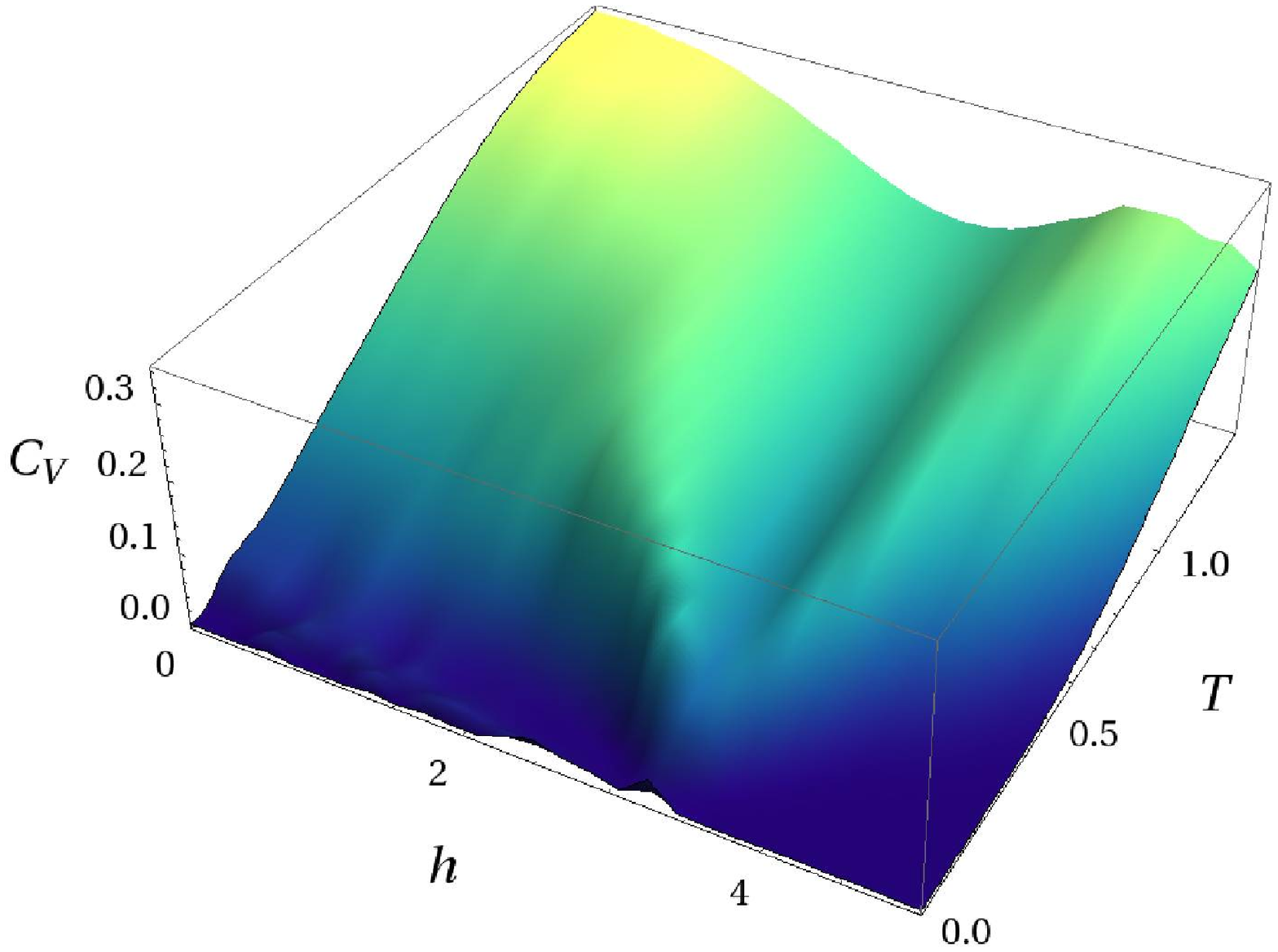}

\includegraphics[width=8cm]{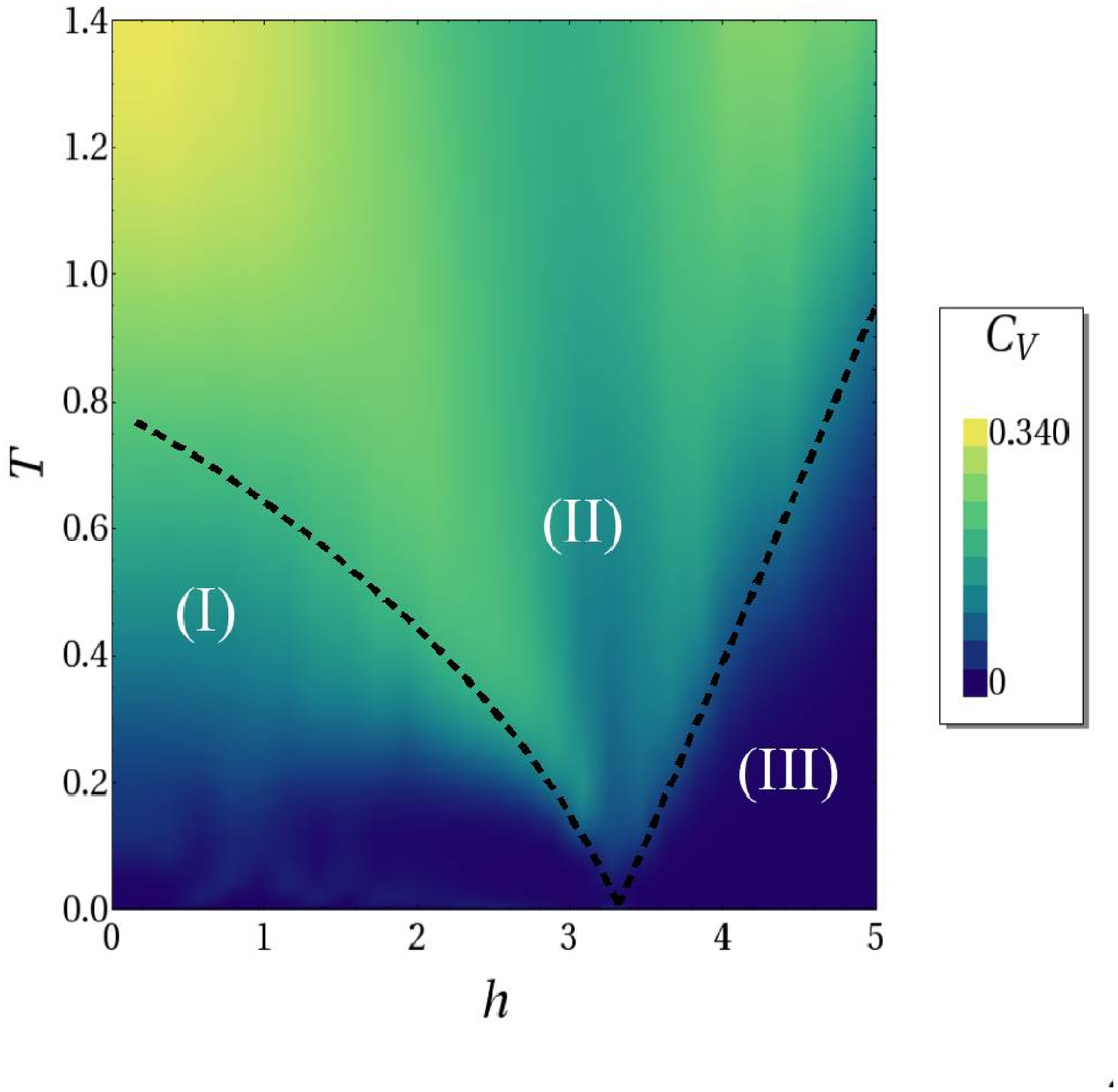}
\caption{The specific heat of XXZ chain in the transverse field ($h$) versus 
temperature ($T$) for $\Delta=0.25$. The data are from LTLM results and $N=20$.
(a) The three dimensional plot, (b) the density plot.(Color online)}
\label{fig2}
\end{center}
\end{figure}

We have plotted the specific heat of the XXZ chain in the transverse field ($h$)
versus temperature ($T$) in Fig. \ref{fig2}. The specific heat is computed using
\begin{equation}
 C_v = \dfrac{J}{k_B T^2} ( \langle E^2 \rangle - \langle E \rangle^2 ).
\end{equation}

The data comes from the Lanczos algorithm with random sampling using $R=100$, $M=30-100$ 
(to get 8 digits accuracy in the first excited state energy) and for
a chain of length $N=20$ with $\Delta=0.25$. For $h<h_c\simeq3.3$ the specific heat
shows exponential decay as $T$ goes to zero. This is in agreement with the presence
of a finite energy gap in this region. The model is gapless at $h=0$, however, our
data on a finite size does not show this behavior because of the finite size effects.
The finite size effects also appear as some level crossing between the first excited
state and the ground state in finite system  which become degenerate ground state in the
thermodynamic limit ($N\rightarrow \infty$). This level crossing causes small oscillations 
close to $T=0$ and for $h<h_c$.
The finite size analysis shows that the gap ($E_g$) scales as $E_g \sim h^{\nu(\Delta)}$ 
where $\nu(\Delta=0.25)=1.18\pm 0.01$ \cite{dmitriev1,dmitriev2,gap}. 
It is the scaling behavior of the second excited state which becomes gapped in the
thermodynamic limit ($N\rightarrow \infty$). Our data confirms the exponential decay
of the specific heat for very low temperatures as $h<h_c$.

 However, the general feature $-$ which is the opening of energy gap due
to breaking of rotational $U(1)$ symmetry $-$ has been appeared as the exponential decay
of specific heat at low temperatures. The exponential decay is vanishing as $h\rightarrow h_c$.
At the critical point ($h=h_c$) the gap vanishes due to the presence of Goldestone modes
which destroy the long-range antiferromagnetic order in $y$-direction. For $h>h_c$ the 
system enters the paramagnetic phase and the gap is proportional to the $h-h_c$ which
is clearly shown in Fig. \ref{fig2}.

Moreover, the density plot in Fig. \ref{fig2} shows 
three different regions at finite temperature which are distinguished by different 
energy scales. (I) The low temperature and $h<h_c$ where the classical domain walls quasi particles 
define the dynamic of system \cite{sachdev1}. (II) The quantum critical region where 
the dynamic is dictated by the quantum fluctuations which become long ranged as both
the temporal and spatial correlation length diverge. (III) The low temperature and
$h>h_c$ where the classical spin flipped quasi particles represent the dynamical behavior 
of the model. 
%%%%%%%%%%%%%%%%%%%%%%%%%%%%%%%%%%%%%%%%%%%%%%%%%%%%%%%%%%%%%%%%%%%%%%%%%%%
\begin{figure}[ht!]
\begin{center}
\includegraphics[width=8cm]{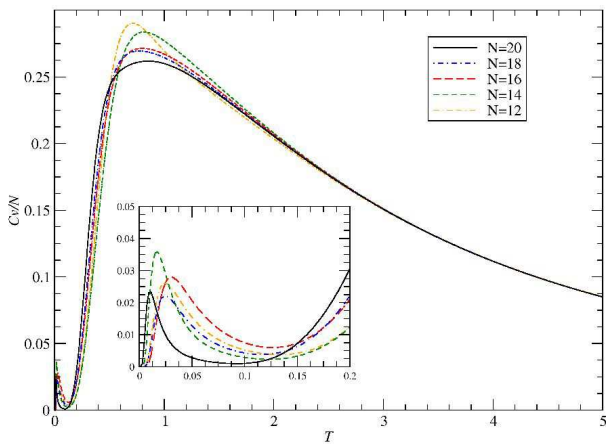}
\caption{The specific heat of XXZ chain versus 
temperature ($T$) in the transverse field $h=1$ and  $\Delta=0$ 
for different chain lengths $N=10, 12, 14,16,18,20$. The inset shows the very
low temperature regime.(Color online)}
\label{fig3}
\end{center}
\end{figure}
%%%%%%%%%%%%%%%%%%%%%%%%%%%%%%%%%%%%%%%%%%%%%%%%%%%%%%%%%%%%%%%%%%%%%%%%%%%

We would like here to comment on the finite size effects in our calculations. Since
the LTLM approach is based on several sampling in the Hilbert space, the finite size effect
will be weak and negligible for nonzero temperature. 
The size dependence of the specific heat is shown in 
Fig. (\ref{fig3}) where different chains with $N=10, 12, 14, 16, 18, 20$ have been considered.
Unless the height of specific heat which is not the same for different $N$, in the
low and high temperature regions all data fall on each other which justifies the 
weak finite size effect. We ignore the small size dependence and claim that the 
our data of the specific heat can be interpreted for a very large chain length.
The inset of Fig. (\ref{fig3}) shows the small oscillations in very low temperature which
is the result of level crossing where the ground state expectation is dominant 
in the partition function and will vanish as the chain size becomes large.

A remark is in order here. The finite size effect is weak and is compensated by random sampling
if the model is not close to a quantum critical point. This is the case of our data presented
in Fig. (\ref{fig3}) at $h=1$. However, close to quantum critical points a finite size analysis 
should be implemented to find the correct behavior. For instance, the level crossing of the
first two excited states should be incorporated in a finite size treatment as done in 
Ref. \onlinecite{gap} to find the scaling of energy gap. The situation is more complex
to find the scaling behavior close to $h_c$ since the value of $h_c$ itself is obtained
by a finite accuracy which is an extra source of error in determing the critical exponents.

We have also calculated the specific heat of the XY chain, $\Delta=0$, which shows similar
qualitative behavior to Fig. \ref{fig2}. It is in agreement with the quantum renormalization group
results \cite{langari} which states that the universality class of $0<\Delta<1$ 
is the same as $\Delta=0$ case. However, the critical field ($h_c$) is slightly smaller
for $\Delta=0$ which is approximately, $h_c(\Delta=0)\simeq3.1$.

\subsection{Static Structure Factors}
We have calculated the static structure factor at finite temperature. 
The $\alpha$-component spin structure factor at momentum $p$ is defined by
\begin{equation}
G^{\alpha \alpha}(p)=\sum_{x=1}^N \langle s_1^{\alpha} s_{1+x}^{\alpha} \rangle e^{ipx},
\label{structurefactor}
\end{equation}
where $\langle \dots \rangle$ is the thermal average defined in Eq. (\ref{ltlm}).
%%%%%%%%%%%%%%%%%%%%%%%%%%%%%%%%%%%%%%%%%%%%%%%%%%%%%%%%%%%%%%%%%%%%%%%%%%
\begin{figure}[ht!]
\begin{center}
\includegraphics[width=8cm]{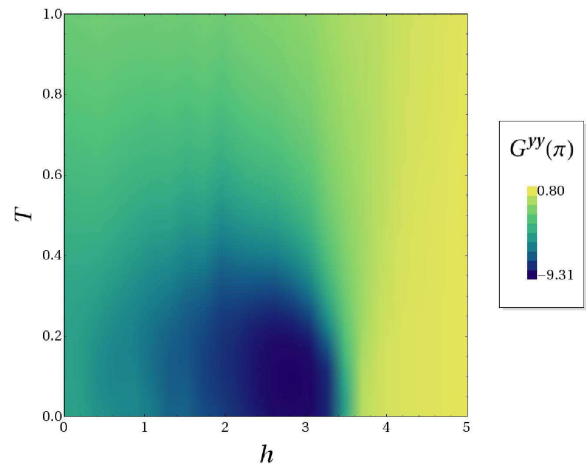}
\caption{The density plot of $y$-component spin structure factor at momentum $p=\pi$
versus the transverse magnetic field ($h$) and temperature ($T$). 
The chain length is $N=20$ and $\Delta=0.25$.(Color online)}
\label{fig4}
\end{center}
\end{figure}

\begin{figure}[ht!]
\begin{center}
\includegraphics[width=8cm]{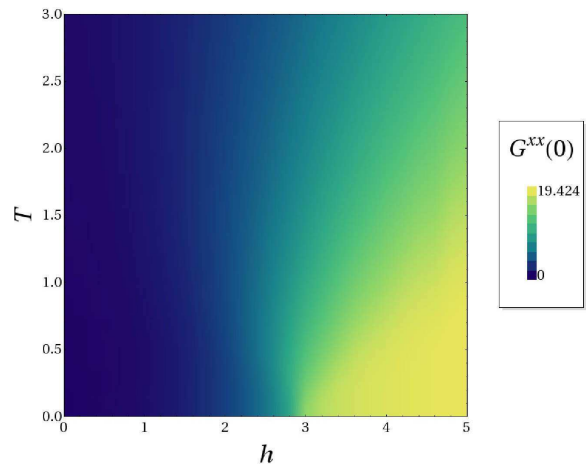}
\caption{The density plot of $x$-component spin structure factor at momentum $p=0$
versus the transverse magnetic field ($h$) and temperature ($T$). 
The chain length is $N=20$ and $\Delta=0.25$.(Color online)}
\label{figC}
\end{center}
\end{figure}

%%%%%%%%%%%%%%%%%%%%%%%%%%%%%%%%%%%%%%%%%%%%%%%%%%%%%%%%%%%%%%%%%%%%%%%%%%

The Hamiltonian (Eq. (\ref{xxztf})) is gapless at $h=0$ where there is no long range ordering
in the ground state while the correlation functions decay algebraically, it is called spin fluid
state. The transverse field ($h\neq0$) breaks the U(1) rotational symmetry to a lower Ising-like
which develops a nonzero energy gap. The ground state then has long range antiferromagnetic order
for $0\leq \Delta <1$. However, due to nonzero projection of the magnetization in the direction
of the transverse field, it is a spin-flop phase. 
%%%%%%%%%%%%%%%%%%%%%%%%%%%%%%%%%%%%%%%%%%%%%%%%%%%%%%%%%%%%%%%%%%%%%%%%%%
\begin{figure}
\begin{center}
\includegraphics[width=8cm]{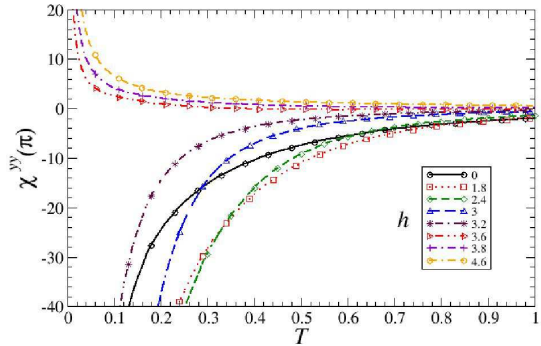}
\caption{The RPA three dimensional $\chi^{yy}(\pi)$ versus temperature for 
different transverse field, $\Delta=0.25$ and $J_{\bot}\simeq 0.00458 J$ for Cs$_2$CoCl$_4$. 
For $h<h_c$ the susceptibility diverges at finite
temperature while for $h>h_c$ it only diverges at $T=0$.(Color online)}
\label{fig5}
\end{center}
\end{figure}
%%%%%%%%%%%%%%%%%%%%%%%%%%%%%%%%%%%%%%%%%%%%%%%%%%%%%%%%%%%%%%%%%%%%%%%%%%%%%%%

The antiferromagnetic ordering in $y$-direction for the spin-flop phase implies 
that the $y$-component structure factor at $p=\pi$ diverges as the size of system 
goes to infinity. This can be seen as a peak in the structure factor.

The density plot of $G^{yy}(\pi)$ is shown in Fig. (\ref{fig4}) versus temperature ($T$)
and transverse field ($h$) for $\Delta=0.25$. The negative peak in $G^{yy}(\pi)$ (which is a signature of 
antiferromagnetic ordering) is the dark (blue) area in Fig. (\ref{fig4}). It shows that the ordering is 
absent at $h=0$ and will gradually appears for increasing $h$. The antiferromagnetic ordering in
$y$-direction can be represented by staggered magnetization in the same direction ($sm_y$) 
as the order parameter. Fig.(\ref{fig4}) shows that $sm_y$ becomes maximum at $h\simeq2$.
The point $h\simeq2$ can also represent the position where the gap becomes maximum. Further
increasing the transverse field suppress the antiferromagnetic ordering until reaching
the critical point $h=h_c$. It is clearly shown in Fig.(\ref{fig4}) that $sm_y$ becomes 
zero at $h=h_c$.

The density plot of x-component structure factor at $p=0$ has been presented in
Fig.(\ref{figC}) versus $T$ and $h$. The value of $G^{xx}(0)$ is small for $h\lesssim 2$
which implies low alignment of the spins in the direction of the external field. However, this value
starts to grow for $h>2$ and obviously saturates for $h \geq h_c=3.3$. The maximum value of
$G^{xx}(0)$ is the signature of long range order in the field direction where the moments
saturates at the maximum value in the paramagnetic phase. However, as the temperature increases the
ordering will be washed out due to thermal fluctuations which destroy the long range 
order in one dimension. The long range order at finite temperature is meaningful in the
sense of three dimensional model which is composed of weakly coupled chains. This will be explained
in the next sections via the random phase approximation (RPA).

\subsection{Magnetic Susceptibility}
Magnetic susceptibility can be calculated using,

\begin{equation}
\chi^\alpha_{1D} \cong \dfrac{1}{T} G^{\alpha\alpha}(k).
\end{equation}

%%%%%%%%%%%%%%%%%%%%%%%%%%%%%%%%%%%%%%%%%%%%%%%%%%%%%%%%%%%%%%%%%%%%%%%%%%
\begin{figure}[ht!]
\begin{center}
\includegraphics[width=8cm]{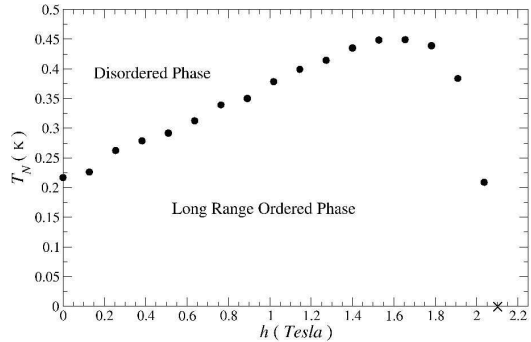}
\caption{The phase diagram of the bulk material in $T$ (temperature) and
$h$ (magnetic field) plane. It is the result of LTLM for one dimensional chain
and random phase approximation (RPA) to get the bulk property with 
inter-chain exchange coupling $J_{\bot}\simeq 0.00458 J$ for Cs$_2$CoCl$_4$.
The cross ($\times$) represents the zero temperature critical field.}
\label{fig6}
\end{center}
\end{figure}
%%%%%%%%%%%%%%%%%%%%%%%%%%%%%%%%%%%%%%%%%%%%%%%%%%%%%%%%%%%%%%%%%%%%%%%%%%%%%%%

To get the susceptibility of the bulk material which is composed of coupled
chains with inter-chain exchange, $J_{\bot}$, we have used the random phase approximation (RPA).
It is a mean field approach where the one dimensional chain is treated exactly while the
interactions between chains are weak and considered in a mean field approach.
In the disordered phase the dynamical susceptibility of coupled chains
at zero frequency is \cite{dmitriev3}
\begin{equation}
\chi^{yy}(p)=\frac{\chi^{yy}_{1D}(p)}{1-J_{\bot} \chi^{yy}_{1D}(p)},
\end{equation}
where $\chi^{yy}_{1D}(p)$ is the one dimensional susceptibility in $y$-direction and momentum $p$
which comes from the LTLM computations. We have plotted $\chi^{yy}(p=\pi)$ versus temperature 
in Fig. (\ref{fig5}) for different transverse field. The susceptibility diverges at finite
(nonzero) temperature for $h<h_c$ which shows the phase transition from the disordered phase
to the antiferromagnetic long range ordered phase. 
For $h>h_c$, the susceptibility only diverges
at $T=0$ which justifies no long range ordered at finite temperature.

We have also used the RPA to calculate the inter-chain 
exchange coupling ($J_{\bot}$) from our numerical computations and experimental data \cite{Kenzelmann}. This can help us to justify the mean field treatment of the 
interactions between chains.
We have used the $h=0$ and $T_N\simeq 217^{mK}\simeq 0.0813 J$ data
presented in Ref. \onlinecite{Kenzelmann,Radu} and solved $1-J_{\bot} \chi^{yy}_{1D}(T_N, \pi)=0$
to find $J_{\bot}$. The result of this calculation is $J_{\bot}\simeq 0.00458 J$ which 
can be compared with $J_{\bot}\simeq 0.0147 J$ estimated in  Ref. \onlinecite{dmitriev3}.
The coupling between chains is two order of magnitude smaller than the intrachain interaction
which verifies the RPA implemented here.
We have then used the value of $J_{\bot}\simeq 0.00458 J$ to find the whole phase diagram 
of the bulk material using our numerical data at finite magnetic field. The phase
diagram of the three dimensional system has been presented in Fig. (\ref{fig6}).
The phase diagram shows the border between the ordered and disordered phase in 
the $T-h$ plane. This is in good agreement with the experimental data presented
in Fig. 11 of Ref. \onlinecite{Kenzelmann}.

\begin{figure}
\includegraphics[width=8cm]{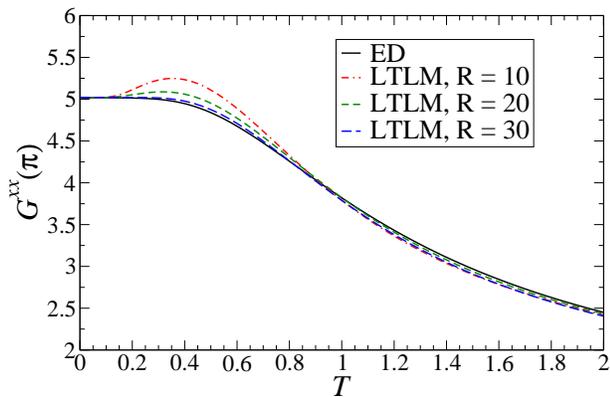}
\caption{The effect of number of random samples in $X$-component of spin structure factor at momentum $p=\pi$ versus temperature, (Color online)}
\label{figA}
\end{figure}

\begin{figure}
\includegraphics[width=6cm]{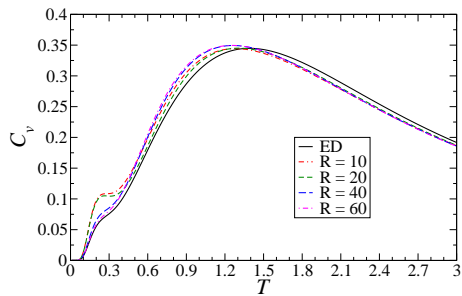}
\caption{Specific heat vs. temperature with different number of random samples, with $h=0$ and $\Delta=0.25$. The solid black line is the result of exact diagonalization. (Color online)}
\label{figB}
\end{figure}

\subsection{Summary and discussion}

We have implemented the symmetric algorithm of the Lanczos method to find the 
finite temperature (thermodynamic) properties of the anisotropic Heisenberg model in the presence
of a transverse magnetic field. It is argued that the symmetric algorithm \cite{aichhorn} (LTLM)
is more accurate at low temperatures than the finite temperature Lanczos method \cite{jaklic}.
We have found that the specific heat decays exponentially at low temperatures 
for $h<h_c$ which justifies the presence of finite energy gap in the lowest part of
spectrum. The gap vanishes both at $h=0$ and $h=h_c$ with a maximum around $h=2$.
A similar behavior has been observed for the y-component structure factor at the
antiferromagnetic wave vector $p=\pi$. It shows that the onset of transverse field 
opens a gap which stabilizes the antiferromagnetic (AF) order in y-direction. 
The AF ordering becomes maximum at $h=2$ and start to decrease by further increasing
of the magnetic field. The long range magnetic order vanishes at the critical point ($h=h_c$)
where the gap becomes zero and the Goldestone mode destroy the ordering. 
The data of the one dimensional model can be used within a random phase approximation
to find the phase diagram of a the experiments \cite{Kenzelmann} done on Cs$_2$CoCl$_4$.
We have estimated the interchain coupling to be $J_{\bot}\simeq 0.00458 J$ which justifies
the RPA method. Moreover, we have plotted the phase diagram of the three dimensional
model (weakly coupled chains) in Fig.(\ref{fig6}) which is in a very good agreement 
with the experimental results presented in Ref.\onlinecite{Kenzelmann}.

We have shown that the LTLM results are not finite size dependent if they are
not close to the quantum critical points. It is the result of random sampling 
which exists in the algorithm. However, a finite size analysis is required to
find the scaling behavior close to quantum critical points. The scaling property of
the energy gap has already been discussed in Ref.\onlinecite{gap}, while there
are other aspects which are open for further investigations.

We would like also to comment on the number of random sampling in this method.
We have plotted the x-component structure factor at $p=\pi$ versus temperature in
Fig.\ref{figA} for different sampling, $R=10, 20, 30$ and the exact diagonalization
on the full spectrum.  It shows that $R=30$ is enough to get a good accuracy for
the finite temperature behavior. A similar data for the specific heat versus
temperature and $R=10, 20, 40, 60$ in Fig.\ref{figB} shows that the number of
$R=60$ sampling reproduces well the exact results. We then conclude that our results
is reliable for $R=100$. One should note that the CPU required time is 
proportional to the number of sampling (R).

\begin{acknowledgments}
The authors would like to thank P. Thalmeier, M. Kohandel, B. Schmidt and H. Rezania for their valuable comments and discussions. This work was made possible by the facilities of the Shared Hierarchical Academic Research
Computing Network (SHARCNET:www.sharcnet.ca). This work was supported in part by the center of excellence in Complex
Systems and Condensed Matter (www.cscm.ir).
\end{acknowledgments}

%\bibliography{FTLM-XXZ.bib}

\begin{thebibliography}{18}
\expandafter\ifx\csname natexlab\endcsname\relax\def\natexlab#1{#1}\fi
\expandafter\ifx\csname bibnamefont\endcsname\relax
  \def\bibnamefont#1{#1}\fi
\expandafter\ifx\csname bibfnamefont\endcsname\relax
  \def\bibfnamefont#1{#1}\fi
\expandafter\ifx\csname citenamefont\endcsname\relax
  \def\citenamefont#1{#1}\fi
\expandafter\ifx\csname url\endcsname\relax
  \def\url#1{\texttt{#1}}\fi
\expandafter\ifx\csname urlprefix\endcsname\relax\def\urlprefix{URL }\fi
\providecommand{\bibinfo}[2]{#2}
\providecommand{\eprint}[2][]{\url{#2}}

\bibitem[{\citenamefont{Sachdev}(1999)}]{Sachdevbook}
\bibinfo{author}{\bibfnamefont{S.}~\bibnamefont{Sachdev}},
  \emph{\bibinfo{title}{{Quantum phase transitions}}}
  (\bibinfo{publisher}{Cambridge University Press}, \bibinfo{year}{1999}).

\bibitem[{\citenamefont{Vojta}(2003)}]{Vojta03}
\bibinfo{author}{\bibfnamefont{M.}~\bibnamefont{Vojta}}, \bibinfo{journal}{Rep.
  Prog. Phys.} \textbf{\bibinfo{volume}{66}} (\bibinfo{year}{2003}).

\bibitem[{\citenamefont{Affleck and Oshikawa}(1999)}]{affleck}
\bibinfo{author}{\bibfnamefont{I.}~\bibnamefont{Affleck}} \bibnamefont{and}
  \bibinfo{author}{\bibfnamefont{M.}~\bibnamefont{Oshikawa}},
  \bibinfo{journal}{Phys. Rev. B} \textbf{\bibinfo{volume}{60}},
  \bibinfo{pages}{1038} (\bibinfo{year}{1999}).

\bibitem[{\citenamefont{Uimin et~al.}(2000)\citenamefont{Uimin, Kudasov, Fulde,
  and Ovchinnikov}}]{uimin}
\bibinfo{author}{\bibfnamefont{G.}~\bibnamefont{Uimin}},
  \bibinfo{author}{\bibfnamefont{Y.}~\bibnamefont{Kudasov}},
  \bibinfo{author}{\bibfnamefont{P.}~\bibnamefont{Fulde}}, \bibnamefont{and}
  \bibinfo{author}{\bibfnamefont{A.}~\bibnamefont{Ovchinnikov}},
  \bibinfo{journal}{Euro. Phys. J. B} \textbf{\bibinfo{volume}{16}},
  \bibinfo{pages}{241} (\bibinfo{year}{2000}).

\bibitem[{\citenamefont{Kenzelmann et~al.}(2002)}]{Kenzelmann}
\bibinfo{author}{\bibfnamefont{M.}~\bibnamefont{Kenzelmann}}
  \bibnamefont{et~al.}, \bibinfo{journal}{Phys. Rev. B}
  \textbf{\bibinfo{volume}{65}}, \bibinfo{pages}{144432}
  (\bibinfo{year}{2002}).

\bibitem[{\citenamefont{Radu}(2002)}]{Radu}
\bibinfo{author}{\bibfnamefont{T.}~\bibnamefont{Radu}}, Ph.D. thesis,
  \bibinfo{school}{Max Planck Institute for Chemical Physics of Solids}
  (\bibinfo{year}{2002}).

\bibitem[{\citenamefont{Kohgi et~al.}(2001)}]{kohgi}
\bibinfo{author}{\bibfnamefont{M.}~\bibnamefont{Kohgi}} \bibnamefont{et~al.},
  \bibinfo{journal}{Phys. Rev. Lett.} \textbf{\bibinfo{volume}{86}},
  \bibinfo{pages}{2439} (\bibinfo{year}{2001}).

\bibitem[{\citenamefont{Dmitriev
  et~al.}(2002{\natexlab{a}})\citenamefont{Dmitriev, Krivnov, Ovchinnikov, and
  Langari}}]{dmitriev1}
\bibinfo{author}{\bibfnamefont{D.~V.} \bibnamefont{Dmitriev}},
  \bibinfo{author}{\bibfnamefont{V.~Y.} \bibnamefont{Krivnov}},
  \bibinfo{author}{\bibfnamefont{A.~A.} \bibnamefont{Ovchinnikov}},
  \bibnamefont{and} \bibinfo{author}{\bibfnamefont{A.}~\bibnamefont{Langari}},
  \bibinfo{journal}{JETP} \textbf{\bibinfo{volume}{95}}, \bibinfo{pages}{538}
  (\bibinfo{year}{2002}{\natexlab{a}}).

\bibitem[{\citenamefont{Dmitriev
  et~al.}(2002{\natexlab{b}})\citenamefont{Dmitriev, Krivnov, and
  Ovchinnikov}}]{dmitriev2}
\bibinfo{author}{\bibfnamefont{D.}~\bibnamefont{Dmitriev}},
  \bibinfo{author}{\bibfnamefont{V.}~\bibnamefont{Krivnov}}, \bibnamefont{and}
  \bibinfo{author}{\bibfnamefont{A.}~\bibnamefont{Ovchinnikov}},
  \bibinfo{journal}{Phys. Rev. B} \textbf{\bibinfo{volume}{65}},
  \bibinfo{pages}{172409} (\bibinfo{year}{2002}{\natexlab{b}}).

\bibitem[{\citenamefont{Dmitriev and Krivnov}(2004)}]{dmitriev3}
\bibinfo{author}{\bibfnamefont{D.}~\bibnamefont{Dmitriev}} \bibnamefont{and}
  \bibinfo{author}{\bibfnamefont{V.}~\bibnamefont{Krivnov}},
  \bibinfo{journal}{Phys. Rev. B} \textbf{\bibinfo{volume}{70}},
  \bibinfo{pages}{144414} (\bibinfo{year}{2004}).

\bibitem[{\citenamefont{Langari}(2004)}]{langari}
\bibinfo{author}{\bibfnamefont{A.}~\bibnamefont{Langari}},
  \bibinfo{journal}{Phys. Rev. B} \textbf{\bibinfo{volume}{69}},
  \bibinfo{pages}{100402} (\bibinfo{year}{2004}).

\bibitem[{\citenamefont{Langari and Mahdavifar}(2006)}]{gap}
\bibinfo{author}{\bibfnamefont{A.}~\bibnamefont{Langari}} \bibnamefont{and}
  \bibinfo{author}{\bibfnamefont{S.}~\bibnamefont{Mahdavifar}},
  \bibinfo{journal}{Phys. Rev. B} \textbf{\bibinfo{volume}{73}},
  \bibinfo{pages}{54410} (\bibinfo{year}{2006}).

\bibitem[{\citenamefont{S. and P.}(1997)}]{sachdev-young}
\bibinfo{author}{\bibfnamefont{S.}~\bibnamefont{S.}} \bibnamefont{and}
  \bibinfo{author}{\bibfnamefont{Y.~A.} \bibnamefont{P.}},
  \bibinfo{journal}{Phys. Rev. Lett.} \textbf{\bibinfo{volume}{78}},
  \bibinfo{pages}{2220} (\bibinfo{year}{1997}).

\bibitem[{\citenamefont{H.~A.~Algra and Carlin}(1976)}]{algra}
\bibinfo{author}{\bibfnamefont{H.~W. J. B. W. J.~H.} \bibnamefont{H.~A.~Algra},
  \bibfnamefont{L.~J. de~Jongh}} \bibnamefont{and}
  \bibinfo{author}{\bibfnamefont{R.~L.} \bibnamefont{Carlin}},
  \bibinfo{journal}{Physica (Utrecht)} \textbf{\bibinfo{volume}{82B}},
  \bibinfo{pages}{239} (\bibinfo{year}{1976}).

\bibitem[{\citenamefont{Jakli{\v{c}} and Prelov{\v{s}}ek}(1994)}]{jaklic}
\bibinfo{author}{\bibfnamefont{J.}~\bibnamefont{Jakli{\v{c}}}}
  \bibnamefont{and}
  \bibinfo{author}{\bibfnamefont{P.}~\bibnamefont{Prelov{\v{s}}ek}},
  \bibinfo{journal}{Phys. Rev. B} \textbf{\bibinfo{volume}{49}},
  \bibinfo{pages}{5065} (\bibinfo{year}{1994}).

\bibitem[{\citenamefont{Aichhorn et~al.}(2003)\citenamefont{Aichhorn, Daghofer,
  Evertz, and von~der Linden}}]{aichhorn}
\bibinfo{author}{\bibfnamefont{M.}~\bibnamefont{Aichhorn}},
  \bibinfo{author}{\bibfnamefont{M.}~\bibnamefont{Daghofer}},
  \bibinfo{author}{\bibfnamefont{H.}~\bibnamefont{Evertz}}, \bibnamefont{and}
  \bibinfo{author}{\bibfnamefont{W.}~\bibnamefont{von~der Linden}},
  \bibinfo{journal}{Phys. Rev. B} \textbf{\bibinfo{volume}{67}},
  \bibinfo{pages}{161103} (\bibinfo{year}{2003}).

\bibitem[{\citenamefont{Siahatgar}(2007)}]{siahatgar}
\bibinfo{author}{\bibfnamefont{M.}~\bibnamefont{Siahatgar}}, Master's thesis,
  \bibinfo{school}{Sharif University of Technology} (\bibinfo{year}{2007}).

\bibitem[{\citenamefont{Sachdev and Young}(1997)}]{sachdev1}
\bibinfo{author}{\bibfnamefont{S.}~\bibnamefont{Sachdev}} \bibnamefont{and}
  \bibinfo{author}{\bibfnamefont{A.}~\bibnamefont{Young}},
  \bibinfo{journal}{Phys. Rev. Lett.} \textbf{\bibinfo{volume}{78}},
  \bibinfo{pages}{2220} (\bibinfo{year}{1997}).

\end{thebibliography}
%\bibliographystyle{apsrev}

\end{document}